\documentclass[aps,prl,preprint,groupedaddress,showpacs]{revtex4}
\usepackage{graphics}
\usepackage{bm}

\begin{document}


\title{Spectroscopic imaging of single atoms within a bulk solid}


\author{M. Varela,$^1$ S. D. Findlay,$^2$ A. R. Lupini,$^1$
H. M. Christen,$^1$ A. Y. Borisevich,$^1$ N. Dellby,$^3$ O. L.
Krivanek,$^3$ P. D. Nellist,$^3$ M. P. Oxley,$^2$ L. J. Allen,$^2$
and S. J. Pennycook$^1$}

\affiliation{$^1$Oak Ridge National Laboratory, Condensed Matter
Sciences Division, P.O. Box 2008, Oak Ridge, TN 37831-6030,
USA\\$^2$School of Physics, University of Melbourne, Victoria
3010, Australia\\$^3$Nion Co., 1102 8th St., Kirkland, WA 98033,
USA}


\begin{abstract}
The ability to localize, identify and measure the electronic
environment of individual atoms will provide fundamental insights
into many issues in materials science, physics and nanotechnology.
We demonstrate, using an aberration-corrected scanning
transmission electron microscope, the spectroscopic imaging of
single La atoms inside CaTiO$_3$. Dynamical simulations confirm
that the spectroscopic information is spatially confined around
the scattering atom. Furthermore we show how the depth of the atom
within the crystal may be estimated.
\end{abstract}


\maketitle


Detection and measurement of the response of individual atoms has
become a challenging issue to provide new insight into many fields
in condensed matter and nanoscale sciences. Distributions of
isolated atoms deeply modify the physical properties of many of
the technologically most relevant and scientifically most
interesting materials. Therefore, analytical techniques capable of
probing single-atom identity and location are increasingly in
demand. Here we show how the aberration-corrected scanning
transmission electron microscope (STEM) allows not only the
imaging of individual atoms inside a crystal, but their
spectroscopic identification, with spatial resolution at the
atomic scale.
This substantial improvement in sensitivity opens up
the possibility of probing the electronic environment of a single
atom. Furthermore, by comparing signals from columns adjacent to
that containing the single atom with dynamical simulations, the
depth of the atom in the crystal can be estimated.

Much of the reported work on the identification and imaging of
single atoms has been achieved by indirect techniques such as
image simulation, including complex reconstruction of phase images
with through focal series restoration \cite{JLU,MEA}. The STEM
provides direct images of individual atoms as first demonstrated
many years ago by the imaging of individual U and Th atoms on a
thin carbon film \cite{CWL}. More recently, at a resolution of
$\sim0.13$ nm, the STEM has successfully imaged single atoms on
and within a variety of materials, including catalysts \cite{NP1}
and semiconductors \cite{CWL,VMG,LP}. Although the optics of the
STEM is ideally suited to simultaneous imaging and spectroscopy,
very few results have been reported that combine direct imaging
with spectroscopic identification of individual atoms. Electron
energy loss spectroscopy (EELS) signals also display high spatial
resolution. Sub-unit cell analysis was first achieved by Spence
and Lynch \cite{SL1} at 1.1 nm resolution. Single Gd atoms were
for the first time identified by Suenaga and coworkers \cite{SEA}
by combining phase contrast TEM imaging with EELS mapping in the
STEM. However, their EELS resolution was only 0.6 nm, sufficient
to resolve their widely spaced Gd atoms but insufficient to
resolve the atomic structure of most materials. Scanning probe
microscopies have also succeeded in achieving atomic-resolution
spectroscopic identification of single atoms on surfaces,
but they cannot probe individual atoms within the bulk
environment.

Recent advances in aberration-corrected STEM provide the perfect
scenario to address such a problem \cite{DKN}. A Z-contrast image
is obtained by collecting, point by point, the
Rutherford-scattered electrons at high angles using a high angle
annular dark field (HAADF) detector. Energy loss electrons passing
through the central hole in the detector can be collected
simultaneously with the HAADF image. This correlation allows EELS
to be performed with atomic resolution \cite{BCP,DBP,KMG}, limited
only by the size of the probe \cite{RP1,AFL}. Recently, by
correcting the aberrations of the STEM probe-forming lens, a
sub-{\AA}ngstrom beam has been demonstrated on a VG Microscopes
HB501UX field emission STEM operating at 120 kV which is equipped
with a Nion aberration corrector \cite{BDK}. In a similar
microscope, operated at 100 kV, a spatial resolution close to 0.11
nm is routinely achieved, and in this work we show how this
enables us not only to image an individual atom within a crystal
but to identify it spectroscopically.

The superlattice sample used for this study was a stack of
CaTiO$_3$ and La$_x$Ca$_{1-x}$TiO$_3$ layers with a very low
concentration of La dopants, prepared by pulsed laser deposition
at $800$ $^{\circ}$C, in an O$_2$ pressure of $75$ mTorr, with a
laser energy of $3$ J/cm$^2$ at $248$ nm and a repetition rate of
$10$ Hz. One commercial CaTiO$_3$ target and a specially prepared
La$_x$Ca$_{1-x}$TiO$_3$ target with $x = 0.04 \pm 0.002$ were
used. The sample is epitaxial and highly ordered, and contains
reference layers of La$_{0.04}$Ca$_{0.96}$TiO$_3$ approximately
2.3 nm thick grown using $100$ shots of the La-doped target.
Layers with correspondingly lower La concentrations were grown by
using fewer shots, and were correspondingly thinner. From this
calibration we calculate that a layer grown using just a single
shot results in a La concentration of 1 atom per $60$ nm$^2$.
Assuming a typical TEM cross-section specimen to be about $10$ nm
thick, on average, $1$ La atom should be found every 16 $\pm$ 4
unit-cell columns. Simple statistical arguments yield a
probability of only 3\% that any given La-containing column
contains more than one La atom.  Such cross-section specimens
were prepared by conventional methods. To minimize surface damage the
final cleaning was performed
by ion milling at 0.5 kV.
Plasma cleaning was also used to prevent sample contamination.

\begin{figure}[ht]
\begin{center}
  \scalebox{0.7}{\includegraphics*{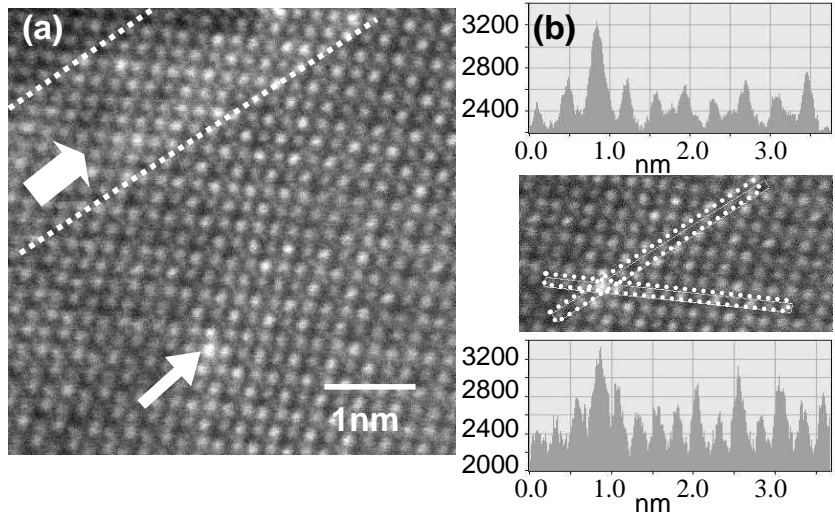}}
\caption{\label{fig:a1} (a) Z-contrast image showing an individual
La atom, marked by a thin arrow. The reference
La$_{0.04}$Ca$_{0.96}$TiO$_3$ layer is marked with a thick arrow
and the interfaces are marked with dotted lines. (b) Intensity
profiles in the [100] and [110] directions (top and bottom
respectively). Column-to-column fluctuations are visible due to
ion milling surface damage, but the La atom is clearly
identifiable and seen to be located on a Ca site. Image obtained
with probe current of approximately 100 pA over 10 s.}
\end{center}
\end{figure}

Figure \ref{fig:a1} shows a high resolution Z-contrast image of a
La$_{0.002}$Ca$_{0.998}$TiO$_3$ layer adjacent to a thicker
La$_{0.04}$Ca$_{0.96}$TiO$_3$ layer. The higher atomic number of
La, as opposed to Ca, O, or Ti, makes single La atoms appear as
bright spots in this environment, although this brightness is
dependent on atom depth \cite{NP2,NKK}. Figure \ref{fig:a1}(b)
shows the intensity traces across the atom. While we cannot
completely exclude the possibility that any particular column
contains two La atoms, many other single La atoms were imaged at
different positions of the sample, in order to get better
statistics, most giving an increase in columnar intensity by $\sim
10\%$. They are seen to sit on the Ca site of the
La$_x$Ca$_{1-x}$TiO$_3$ perovskite structure, easily identifiable
due to the fact that Ca columns appear darker than TiO columns.

\begin{figure}[ht]
\begin{center}
  \scalebox{0.7}{\includegraphics*{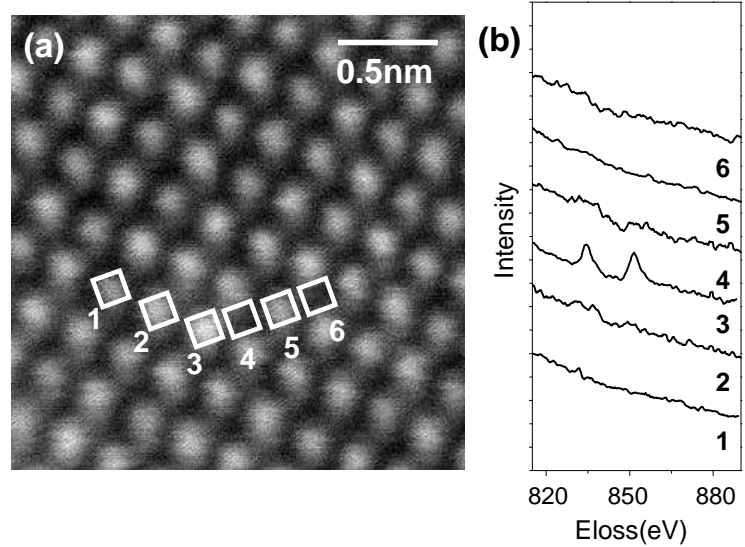}}
\caption{\label{fig:a2}(a) Z-contrast image with (b) EELS traces
showing spectroscopic identification of a single La atom at atomic
spatial resolution, with the same beam used for imaging.  The
M$_{4,5}$ lines of La are seen strongly in spectrum 3 obtained
from the bright column at 20 million magnification and a total
collection time of 30 s.  Other spectra from neighboring columns
show much reduced or undetectable La signal. These spectra were
obtained with collection times of 20 s, and are shown normalized
to the pre-edge intensity and displaced vertically for clarity.}
\end{center}
\end{figure}

Identification of the atoms was carried out using a McMullan
design charge-coupled-device parallel-detection EELS system
\cite{MRM}. This system does not provide optimized coupling from
the aberration-corrected beam, having a collection efficiency of
only 8\%. [Collection efficiency is calculated as $\beta^2 /
(\alpha^2+\theta_{\rm E}^2)$ where $\beta$ = 7 mrad is the
spectrometer entrance aperture, $\alpha$ = 25 mrad is the probe
forming aperture and $\theta_E = \Delta E/2E_0$ is the
characteristic angle of inelastic scattering for an energy loss
$\Delta E$ and incident energy E$_0$.] Nevertheless,  by scanning
the STEM probe directly over a highly magnified image of an
isolated La atom (0.1 nm $\times$ 0.1 nm in size as shown in
figure 2), a clear EELS signature is obtained. In figure
\ref{fig:a2}, spectrum 3 shows two distinct peaks that correspond
to the La M$_{4,5}$ lines at 832 and 849 eV. The intensity is low,
as one would expect from the small number of counts associated
with the excitations of a single atom, but clearly observable. The
total number of M$_{4,5}$ excitations for this spectrum is of the
order of 10$^4$. To be certain that a single La atom can be
detected, ten different bright spots were analyzed and each showed
characteristic La lines with similar intensity. No meaningful beam
damage was observed during the acquisition times employed, which
were of the order of tenths of seconds.

There has been much discussion about detection
limits and the role that the delocalization of the underlying
ionization process plays in limiting the resolution of the STEM
image \cite{SL1,RP1,AFL,KR1,MS1,HB1}. To analyze the
localization of the EELS signal we have taken spectra with the
probe positioned on nearby columns as indicated on figure
\ref{fig:a2}(a). Given the fact that neighboring atomic columns (2
and 4) correspond to different atomic species (TiO and O columns)
the channeling and therefore the localization of the signal would
be different. As depicted in figure \ref{fig:a2}(b), the EELS
signal attributed to the La single atom is clearly localized on
the atomic column containing this ion. When the electron beam is
placed on adjacent columns the signal is substantially reduced, to
only $10\pm5$\% on the neighboring TiO column (labeled 2), and to
$20\pm5$\% on the neighboring O site (labeled 4), only 0.19 nm
away. This clearly indicates that the La can be localized within
significantly better than 0.2 nm, which is much larger than the
impact parameter for La-M excitation \cite{RP1} and therefore
suggests that the residual intensity from neighboring columns is
almost certainly the effect of channeling \cite{AFOR}.

\begin{figure}[h!]
\begin{center}
  \scalebox{0.7}{\includegraphics*{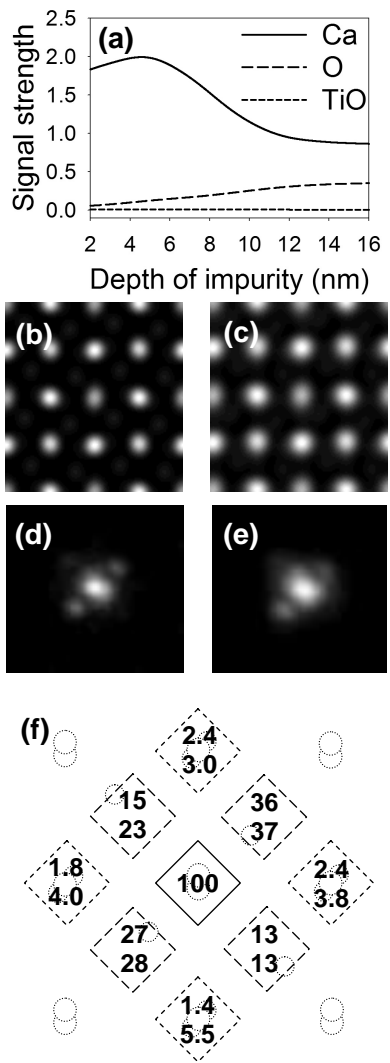}}
\caption{\label{fig:a3} Simulated HAADF and EELS images. (a) Depth
dependence of La signal strength on the different columns using
simple cubic perovskite structure. (b) HAADF image with
aberration-free probe. (c) HAADF image with aberration-balanced
probe and 0.04 nm Gaussian broadening. (d) EELS image
corresponding to probe of (b). (e) EELS image corresponding to
probe of (c). (f) Signal strength in raster scan regions as a
percentage of that on central Ca column. Top number refers to
aberration-free probe. Bottom number refers to aberration-balanced
probe. The boundaries of the scan regions refer to the legend in
(a).}
\end{center}
\end{figure}

To explore the role of channeling, dynamical diffraction
calculations were performed. Figure \ref{fig:a3}(a) shows the
simulated La signal strength for an aberration-free probe situated
on the Ca column, O column, and TiO column as a function of the
impurity depth.  A simplified cubic model for the structure
suffices for this estimate, but subsequent calculations are based
upon the distorted perovskite structure \cite{BM}.  The
contribution on the Ca column initially increases due to the
electrostatic attraction to the column (channeling), but then
decreases as elastic and inelastic scattering mechanisms broaden
the probe. The contribution on the O column rises gradually;
simulations of the electron density make clear that this is due to
probe spreading. From this figure it is estimated that a
$20\pm5\%$ signal on the O column is obtained with an impurity
depth of around 10 nm. The lack of any appreciable signal from the
TiO column is an expression of the channeling strength of this
column. To explain the $10\pm5$\% signal measured it must be
appreciated that in the experiment the probe was not situated upon
a single point but raster scanned across a highly magnified image.
This introduces a finite, square scan region, of approximately 0.1
nm side length. Figures \ref{fig:a3}(b) and \ref{fig:a3}(c) show
the HAADF images simulated \cite{AFOR,FAOR} for a 15 nm thick
crystal using an aberration-free probe and an aberration-balanced
probe respectively.  The latter uses parameters measured from
autotuning which, though recorded on a different occasion, can be
regarded as typical.  The latter image has been convolved with a
Gaussian of half-width 0.04 nm, accounting for position
instability of the probe and finite source size \cite{NR}. EELS
images were simulated using mixed dynamic form factor theory
\cite{FAOR,OA3}.

Figures \ref{fig:a3}(d) and \ref{fig:a3}(e) show EELS images
corresponding to the parameters used in figures \ref{fig:a3}(b)
and \ref{fig:a3}(c). The asymmetry of the structure, particularly
whereby O columns distorted towards the Ca column provide a
stronger signal than those distorted away from it, is evident.
Figure \ref{fig:a3}(f) shows the scan regions and the signal
strengths therein as a percentage of that on the Ca column.  In
these simulations the impurity was set at a depth of 12.2 nm. Note
that the O column signal strength is on average higher than that
of the experiment, and the TiO column signal strength is on
average lower.  For the second probe there are columns which agree
with the experimental results to within the error bars.  While the
parameters used may not correspond precisely to those
characterizing the experiment, the simulations have elucidated the
mechanism by which the signals are obtained. The simulations
support the localization of the ionization interaction and
demonstrate that the dynamical propagation, in transferring some
electron intensity to adjacent columns, is responsible for the
signals obtained.  The depth dependence seen in figure
\ref{fig:a3}(a) is suggestive. It will be interesting to
investigate how accurately the depth of an impurity atom may be
determined in this fashion, particularly using larger
probe-forming apertures and variable defocus.

In summary, an individual atom within a bulk solid has been
identified spectroscopically with spatial resolution at the atomic
level. This not only allows the detection and identification of
single atoms, but also measurement of their electronic properties.
Through dynamical simulations of probe spreading, an indication of
atom depth is also obtained. Furthermore, the local formal
oxidation state can be determined from the ratio of white line
features present in L edges giving direct information about the
magnetic moment and the valence of the atom. The present
statistics are sufficient for such an analysis, although in this
case the La ion is known to have a fixed 3+ valence state.
However, optimum coupling of the aberration-corrected beam into
the spectrometer would not only improve collection efficiency by
over an order of magnitude, but also allow correction of
spectrometer aberrations to give improved energy resolution
\cite{DKM}. This would allow better comparison of local electronic
structure with first-principles calculations \cite{BDP}.

There are many situations where the valence state of localized
impurity atoms is variable, and it is very often the critical
factor controlling macroscopic properties. Examples include
band-bending effects at grain boundaries \cite{KEA,HM}, charge
ordering in perovskite-based oxides \cite{RAK}, charge-modulation
superlattices \cite{OMG}, and molecular species in nanotubes
\cite{SLA,HS1}. We anticipate single atom analysis will open up
many other areas of science where isolated atoms control
macroscopic phenomena, including the origin of ductility or
embrittlement in structural alloys, the nature of trap states at
semiconductor interfaces, luminescent quantum efficiency in
nanocrystals and active sites in heterogeneous catalysts.

The single atom represents the smallest quantum of matter that
retains the characteristics of the material through its local
electronic environment.  The ability to probe the electronic
structure of a material at this level represents the ultimate
advance for understanding the atomic origins of materials
properties. This work opens up this possibility.

M. Varela acknowledges fruitful discussions with R. Sanchez.
Research of M. Varela is performed as a Eugene P. Wigner Fellow
and staff member at ORNL. This research was sponsored by the
Laboratory Directed Research and Development Program of ORNL,
managed by UT-Battelle, LLC, for the U.S. Department of Energy
under Contract No. DE-AC05-00OR22725 and by appointment to the
ORNL Postdoctoral Research Program administered jointly by ORNL
and ORISE. M. P. Oxley and L. J. Allen acknowledge support by the
Australian Research Council.

\bibliography{Varela}

\end{document}